\documentclass[aps,prl,reprint,preprintnumbers,superscriptaddress,amsmath,amssymb,bibnotes,longbibliography]{revtex4-2}
\usepackage{lineno}
\usepackage{graphicx}% Include figure files
\usepackage{bm}% bold math
\usepackage[colorlinks,linkcolor=blue,anchorcolor=blue,citecolor=blue,urlcolor=blue,filecolor=blue,menucolor=blue,runcolor=blue]{hyperref}% add hypertext capabilities
\usepackage{siunitx} %To use units properly
\usepackage[version=4]{mhchem} %For using chemical formulas
\graphicspath{{../}}

\begin{document}

	\title{Pressure-dependent magnetism of the Kitaev candidate Li$_2$RhO$_3$}
	
	\author{Bin Shen}
	\email{bin.shen@physik.uni-augsburg.de}
	\affiliation{Experimental Physics VI, Center for Electronic Correlations and Magnetism, University of Augsburg, 86159 Augsburg, Germany}
	
	\author{Efrain Insuasti Pazmino}
	\affiliation{Felix Bloch Institute for Solid-State Physics, University of Leipzig, 04103 Leipzig, Germany}
	
	\author{Ramesh Dhakal}
	\affiliation{Department of Physics and Center for Functional Materials, Wake Forest University, Winston-Salem, North Carolina 27109, USA}
	
	\author{Friedrich Freund}
	\affiliation{Experimental Physics VI, Center for Electronic Correlations and Magnetism, University of Augsburg, 86159 Augsburg, Germany}
	
	\author{Philipp Gegenwart}
	\affiliation{Experimental Physics VI, Center for Electronic Correlations and Magnetism, University of Augsburg, 86159 Augsburg, Germany}
	
	\author{Stephen M. Winter}
	\affiliation{Department of Physics and Center for Functional Materials, Wake Forest University, Winston-Salem, North Carolina 27109, USA}
	
	\author{Alexander A. Tsirlin}
	\email{altsirlin@gmail.com}
	\affiliation{Felix Bloch Institute for Solid-State Physics, University of Leipzig, 04103 Leipzig, Germany}

	\date{\today}% It is always \today, today,

	\begin{abstract}
		We use magnetization measurements under pressure along with \textit{ab initio} and cluster many-body calculations to investigate magnetism of the Kitaev candidate Li$_2$RhO$_3$. Hydrostatic compression leads to a decrease in the magnitude of the nearest-neighbor ferromagnetic Kitaev coupling $K_1$ and the corresponding increase in the off-diagonal anisotropy $\Gamma_1$, whereas the experimental Curie-Weiss temperature changes from negative to positive with the slope of +40~K/GPa. On the other hand, spin freezing persists up to at least 3.46~GPa with the almost constant freezing temperature of 5~K that does not follow the large changes in the exchange couplings and indicates the likely extrinsic origin of spin freezing. Magnetic frustration in Li$_2$RhO$_3$ is mainly related to the interplay between ferromagnetic $K_1$ and antiferromagnetic $\Gamma_1$, along with the weakness of the third-neighbor coupling $J_3$ that would otherwise stabilize zigzag order. The small $J_3$ distinguishes Li$_2$RhO$_3$ from other Kitaev candidates.

	\end{abstract}
	
	\maketitle
	
	%\tableofcontents
	
	Honeycomb magnets with dominant Kitaev interactions are predicted to realize a spin-liquid state with emergent topological order and exotic excitations~\cite{06Kitaev,trebst2022}. Material realizations of this scenario are usually searched for among the low-spin $d^5$ compounds following the initial proposal by Jackeli and Khaliullin~\cite{09JackeliPRL}. Whereas several honeycomb iridates and Ru$^{3+}$ halides have been extensively studied experimentally~\cite{winter2017, 19JackeliNRP,imai2022}, rhodates remain relatively less explored despite the fact that Rh$^{4+}$ is isoelectronic to Ir$^{4+}$ and well suited for realizing Kitaev exchange. 
	
	Lithium rhodate, Li$_2$RhO$_3$, features a slightly deformed honeycomb lattice of the Rh$^{4+}$ ions~\cite{11TodZAAC} (see figure~\ref{CrisSt}). Its silver analog, Ag$_3$LiRh$_2$O$_6$, can be prepared by an ion-exchange reaction~\cite{bahrami2022}. Despite the rather similar structures of the honeycomb layer, these two compounds feature very different magnetic properties. Whereas the electronic state of Rh$^{4+}$ in Li$_2$RhO$_3$ should be close to $j_{\rm eff}=\frac12$, akin to the typical Ir$^{4+}$ compounds~\cite{winter2017}, a departure from the $j_{\rm eff}=\frac12$ state has been detected in Ag$_3$LiRh$_2$O$_6$ by x-ray spectroscopy~\cite{bahrami2022}. Li$_2$RhO$_3$ evades long-range magnetic order, but reveals a change in spin dynamics associated with spin freezing below 6~K~\cite{13LuoPRB,mazin2013}, as confirmed by nuclear magnetic resonance and muon spectroscopies~\cite{17KhuPRB}. By contrast, Ag$_3$LiRh$_2$O$_6$ develops long-range antiferromagnetic order below 90\,K~\cite{bahrami2022}, which is the highest N\'eel temperature among the $d^5$ honeycomb magnets reported to date. The differences between Li$_2$RhO$_3$ and Ag$_3$LiRh$_2$O$_6$ are far more drastic than between the corresponding iridates, $\alpha$-Li$_2$IrO$_3$ and Ag$_3$LiIr$_2$O$_6$, that reveal rather similar magnetic behavior with the long-range magnetic order below 15\,K~\cite{freund2016} and 8\,K~\cite{bahrami2021}, respectively. This comparison suggests that rhodates may be more tunable by (hydrostatic or chemical) pressure compared to the iridates.
	
	In the following, we explore this possibility and study pressure evolution of Li$_2$RhO$_3$. X-ray diffraction experiments on this compound revealed the structural phase transition with the formation of linear Rh$^{4+}$ chains above 6.5\,GPa at room temperature~\cite{19HerPRB}. It means that Li$_2$RhO$_3$ offers a broader pressure window for tuning magnetism of the Rh$^{4+}$ honeycombs than different polymorphs of Li$_2$IrO$_3$~\cite{tsirlin2022} that become nonmagnetic upon the structural dimerization transition already at $3.5-4.0$\,GPa at room temperature~\cite{hermann2018,majumder2018,clancy2018,hermann2019b,li2020} and at even lower pressures of $1.0-1.5$\,GPa on cooling~\cite{veiga2019,Bin2022}. Here, we probe Li$_2$RhO$_3$ using magnetization measurements under pressure and follow the evolution of individual magnetic couplings in this material using \textit{ab initio} and cluster many-body calculations. Previous quantum-chemistry studies reported Kitaev exchange as one of the leading terms in the Li$_2$RhO$_3$ spin Hamiltonian at ambient pressure~\cite{15KatSR}.

	\section{Results}
	\subsection{Magnetization under pressure}
	
	Figure~\ref{Fig1} (a) shows the inverse magnetic susceptibility $H/M$ of Li$_2$RhO$_3$ measured as a function of temperature under various pressures. A temperature-independent term $\chi_0$, which stands for the residual part of the background from the pressure cell, has been subtracted for each pressure, respectively. At all pressures, $M/H$ monotonically increases upon cooling, similar to the ambient-pressure behavior reported in the literature~\cite{13LuoPRB, 17KhuPRB}. The high-temperature part of magnetic susceptibility can be fitted with the Curie-Weiss law (solid lines), $\chi-\chi_0 = C/(T-\theta)$, where $C$ is the Curie constant and $\theta$ is the Curie-Weiss temperature.
	
	From the fits to the data between 150\,K and 300\,K, we find that the Curie constant and the associated paramagnetic effective moment weakly decrease with pressure, whereas the Curie-Weiss temperature increases and even changes sign from negative to positive (Fig.~\ref{Fig1}), indicating the growth of ferromagnetic interactions. At ambient pressure, the paramagnetic effective moment is around 2.4\,$\mu_B$ in agreement with the previous studies~\cite{17KhuPRB}. It decreases with pressure and at 2\,GPa reaches 1.8\,$\mu_B$, which is close to 1.73\,$\mu_B$ expected for the $j_{\rm eff}=\frac12$ state of Rh$^{4+}$.

	At low temperatures, the smooth evolution of the magnetic susceptibility is consistent with the absence of any magnetic transition. The spin freezing reported below 6\,K at ambient pressure~\cite{13LuoPRB,17KhuPRB} can be tracked by the bifurcation of the magnetic susceptibilities measured under field-cooled (FC) and zero-field-cooled (ZFC) conditions in low applied fields. Such measurements become quite challenging under pressure because of the weak signal in low magnetic fields.  Nevertheless, it was possible to track the FC/ZFC susceptibilities of Li$_2$RhO$_3$ in the applied field of 0.1\,T up to 3.46\,GPa (Figure~\ref{Fig2}). The spin-glass like behavior of Li$_2$RhO$_3$ remains almost unchanged upon compression. The bifurcation of the FC/ZFC susceptibilities is seen at around 5.0~K at all pressures. 
	
	\subsection{Microscopic magnetic model}

	Figure~\ref{PBE+SO} shows the density of states for Li$_2$RhO$_3$ calculated on the full-relativistic (PBE+SO) level. Typically for a transition-metal oxide, the bands near the Fermi level are dominated by the $d$-states that are split into the $t_{2g}$ and $e_g$ manifolds by the octahedral crystal field. The $t_{2g}-e_g$ splitting is about 3.0\,eV compared to the splitting of 2.0\,eV in $\alpha$-RuBr$_3$~\cite{shen2024} and 2.2\,eV in $\alpha$-RuCl$_3$~\cite{majumder2015}, in agreement with the higher negative charge of O$^{2-}$. 
	
	From the orbital energies obtained via the Wannier fit, we extract a small noncubic crystal-field splitting of only 13\,meV, which is well below the spin-orbit coupling constant for Rh$^{4+}$. However, the $t_{2g}$ states do not show the splitting into $j_{\rm eff}=\frac32$ and $j_{\rm eff}=\frac12$ bands known from the Ir$^{4+}$ compounds. There is instead a three-peak structure reminiscent of Na$_2$IrO$_3$~\cite{mazin2012}. This splitting into three sub-bands is a fingerprint of the dominant off-diagonal hopping ($t_2$) that takes place between, e.g., the $d_{yz}$ and $d_{xz}$ orbitals and gives rise to the large Kitaev coupling~\cite{rau2014}. Our direct calculation of the exchange tensor confirms this assessment. 
	
	We define the exchange tensors in the Kitaev coordinate frame as
	\begin{equation}
		\mathbb J=
		\left(
		\begin{array}{ccc}
			J & \Gamma & \Gamma' \\
			\Gamma & J & \Gamma' \\
			\Gamma' & \Gamma' & J+K
		\end{array}
		\right)
		\label{eq:j}\end{equation}
	with four independent parameters ($J$, $K$, $\Gamma$, and $\Gamma'$). Technically, the $X$- and $Y$- bonds have a lower symmetry with six independent parameters~\cite{winter2016}, but our calculations show that the approximate form given by Eq.~\eqref{eq:j} is sufficiently accurate for these bonds as well, so it will be used in the following. The $X$- and $Y$- bonds are symmetry-equivalent but different from the $Z$-bonds. This difference is also relatively small at all pressures, as can be seen in the Supplementary Information. 
	Therefore, in the rest of this work we will discuss the couplings averaged over the $X$-, $Y$-, and $Z$-bonds of the honeycomb lattice. We label these averaged couplings as $\bar J$, $\bar K$, $\bar\Gamma$, etc.

	Figure~\ref{JKGLRO} shows such averaged exchange couplings as a function of pressure. Li$_2$RhO$_3$ is dominated by the ferromagnetic nearest-neighbor Kitaev term $K_1$ that decreases in magnitude upon compression. The off-diagonal anisotropy $\Gamma_1$ increases under pressure, with the $\bar\Gamma_1/|\bar K_1|$ ratio changing from 0.34 at 0\,GPa to 0.60 at 4.5\,GPa. The nearest-neighbor Heisenberg coupling is weakly ferromagnetic and also increases in magnitude upon compression. 
	
	These pressure-induced changes in the nearest-neighbor couplings are well in line with the structural changes upon compression. Indeed, the Rh--O--Rh bond angles systematically decrease, leading to a reduction in the off-diagonal hopping $t_2$ and the weakening of the Kitaev term relative to the other terms of the exchange tensor~\cite{winter2016}. This mechanism appears to be generic for the Kitaev magnets that all show the reduction in the bond angles under hydrostatic pressure and the gradual suppression of the Kitaev term~\cite{majumder2018,Bin2022,shen2024}. The evolution of Li$_2$RhO$_3$ can also be followed on the phase diagram of the $J_1-K_1-\Gamma_1$ model (Figure~\ref{classPD}) where pressure systematically shifts the system away from the Kitaev limit located at $\varphi=3\pi/2$.

	Turning to the smaller terms in the spin Hamiltonian, we note that the off-diagonal anisotropy $\Gamma'$ is below 1\,meV at all pressures. Its negative sign should increase the proclivity of Li$_2$RhO$_3$ for zigzag order, as shown in the phase diagram of Figure~\ref{classPD}. However, the main term stabilizing the zigzag order is believed to be the antiferromagnetic third-neighbor coupling $J_3$ that has been estimated at about $1-3$\,meV in $\alpha$-RuCl$_3$~\cite{winter2017,maksimov2020,laurell2020,samarakoon2022} and $2-6$\,meV in Na$_2$IrO$_3$~\cite{choi2012,katukuri2014,winter2016,kim2020}.  Surprisingly, the $\bar J_3$ of Li$_2$RhO$_3$ is quite small, about 0.15\,meV according to our cluster many-body calculations. This suppression of $\bar J_3$ may be a result of the smaller spatial extent of the Rh $4d$ orbitals compared to the Ir $5d$ orbitals of the iridates, and of the O $2p$ orbitals compared to the Cl $3p$ orbitals of $\alpha$-RuCl$_3$. Interestingly, $\bar J_3$ obtained by the cluster many-body calculations is much lower than in the superexchange model. It means that the hoppings to the $e_g$ orbitals yield ferromagnetic contributions that are strong enough to compensate for antiferromagnetic contributions from the intra-$t_{2g}$ hoppings.

	\section{Discussion}
	
	Li$_2$RhO$_3$ remains a forgotten sibling of the much better known Kitaev iridates and Ru$^{3+}$ halides. Although the first reports of its magnetic properties~\cite{13LuoPRB,mazin2013} even preceded the discovery of $\alpha$-RuCl$_3$ as a Kitaev candidate, relatively little is known about its microscopic regime. The quantum-chemistry study of Li$_2$RhO$_3$ demonstrated a strong Kitaev coupling but also reported an unusually strong spatial anisotropy with $J_1^Z=-10.2$\,meV having a different sign than $J_1^X=2.4$\,meV~\cite{15KatSR}. 
	
	Our study advocates a more conventional microscopic scenario where spatial anisotropy plays only a minor role, and the couplings on the $X/Y$- and $Z$-bonds are qualitatively and quantitatively similar to each other. Moreover, the position of Li$_2$RhO$_3$ on the phase diagram of the extended Kitaev ($J_1-K_1-\Gamma_1$) model should resemble that of the Ru$^{3+}$ halides, with the leading ferromagnetic Kitaev term $K_1<0$ and the subleading off-diagonal anisotropy $\Gamma_1>0$. The nearest-neighbor Heisenberg exchange, $J_1<0$, is relatively small at ambient pressure but becomes increasingly more important upon compression. This trend is corroborated by our magnetization measurements. Indeed, the powder-averaged Curie-Weiss temperature can be calculated as~\cite{winter2016}
	\begin{equation}
		\theta=-(3\bar J_1+\bar K_1)/4k_B
	\end{equation}
	and does not depend on $\bar\Gamma_1$. Whereas $\bar K_1$ decreases in magnitude with pressure, the leading trend is determined by the enhancement of ferromagnetic $\bar J_1$ that causes the increase in $\theta$ upon compression (Fig.~\ref{JKGLRO}(g)). Despite this good qualitative agreement, we note that the slope of the calculated pressure dependence is much lower compared to the experiment. Similar discrepancies have been reported in other Kitaev materials and ascribed to deviations from the simple Curie-Weiss law caused by the temperature dependence of the paramagnetic effective moment~\cite{li2021}.
	
	Although Li$_2$RhO$_3$ does not approach the Kitaev limit, the combination of $\bar K_1<0$ and $\bar\Gamma_1>0$ as the dominant terms in the spin Hamiltonian is a precondition for the strong frustration that has been comprehensively studied in the context of the $K_1-\Gamma_1$ model on the honeycomb lattice~\cite{rousochatzakis2024}. This model connects the limits of the Kitaev spin liquid at large $K_1$ and classical spin liquid at large $\Gamma_1$, whereas the intermediate region is often described as correlated paramagnet~\cite{rousochatzakis2024}. With Li$_2$RhO$_3$ lying close to the $K_1-\Gamma_1$ line on the phase diagram of the $J_1-K_1-\Gamma_1-\Gamma_1'$ model, at least at ambient pressure, it is not surprising that this material evades long-range magnetic order. Experimentally, we find spin freezing below 5\,K that persists at least up to 3.46\,GPa. Interestingly, the freezing temperature almost does not change, whereas individual exchange couplings are clearly affected by pressure. This observation suggests that spin freezing is driven by an extrinsic energy scale and may be associated with the structural disorder arising from stacking faults~\cite{hermann2018,mazin2013}. The high-pressure x-ray diffraction study of Li$_2$RhO$_3$ shows that the concentration of stacking faults is almost unchanged within the pressure range of our study~\cite{19HerPRB}.
	
	Although stacking faults are unavoidable in almost all Kitaev candidates because of their layered nature~\cite{freund2016,bahrami2021,sears2023,zhang2024}, this structural disorder will usually not preclude magnetic ordering. Na$_2$IrO$_3$, $\alpha$-RuCl$_3$, and $\alpha$-RuBr$_3$ all show zigzag magnetic order, which is stabilized by $\Gamma_1'$ and $J_3$. Whereas $\Gamma_1'$ is a comparably minor term across the whole family of the existing Kitaev materials, Li$_2$RhO$_3$ stands apart from the other Kitaev candidates in that its $J_3$ is unusually small. One can then interpret the absence of magnetic order in Li$_2$RhO$_3$ as the joint effect of the frustration caused by $K_1-\Gamma_1$ and the weakness of $J_3$. Additionally, the existing Li$_2$RhO$_3$ samples~\cite{19HerPRB} show about twice higher concentration of stacking faults compared to Na$_2$IrO$_3$~\cite{choi2012}. This increased amount of structural disorder should increase the proclivity for spin freezing.
	
	Finally, we note that our study supports the general trend of tuning Kitaev candidates away from the Kitaev limit by hydrostatic pressure~\cite{tsirlin2022}. One would then expect negative pressure to enhance the Kitaev term, reduce $\Gamma_1/|K_1|$, and bring the materials closer to the Kitaev limit. In this context, it is somewhat surprising that Ag$_3$LiRh$_2$O$_6$, the expanded version of the Li$_2$RhO$_3$ structure, not only shows magnetic ordering, but also features the highest N\'eel temperature among all Kitaev candidates reported to date~\cite{bahrami2022}. The negative pressure effects in honeycomb rhodates may be nontrivial and clearly deserve a further dedicated investigation.
	
	%-------------------------------------------------------------------------------------------------------         
	\section{Methods}
	\subsection{Sample synthesis and characterization}
	Polycrystalline samples of Li$_2$RhO$_3$ used in this work were previously characterized in Refs.~\cite{17KhuPRB, 19HerPRB}. Magnetization under pressure was measured with the same method as in Ref.~\cite{21Bin} using the gasket with the sample chamber diameter of 0.9~mm that can reach pressures up to 2\,GPa. In order to reach higher pressures, the gasket was pre-indented, and pressures up to 3.46\,GPa could be reached in run No. 3. A piece of Pb served to determine pressure from the temperature of its superconducting transition. Daphne oil 7373 was used as pressure-transmitting medium. 
	
	\subsection{DFT and cluster many-body calculations}
\begin{figure*}
	\includegraphics[angle=0,width=1\textwidth]{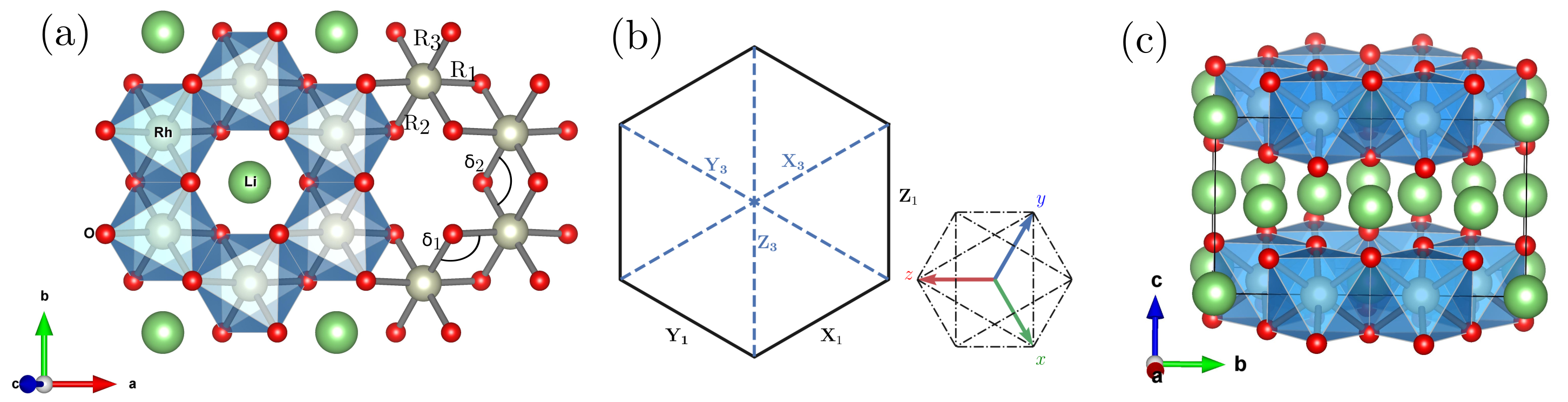}
	\vspace{-12pt} \caption{\label{CrisSt} Structural details of \ce{Li2RhO3}. (a) \ce{Rh}-honeycomb layer along the $ab$-plane showing the two symmetry allowed \ce{Rh-O-Rh} angles and the definition of the \ce{Rh-O} bonds. (b) Definition of the local coordinate reference frame $x,y,z$, and the bonds of interest for first and third neighbors. (c) Stacking of honeycomb layers in the \ce{Li2RhO3} structure.
	}
	\vspace{-12pt}
\end{figure*}

\begin{figure*}
	\includegraphics[angle=0,width=0.8\textwidth]{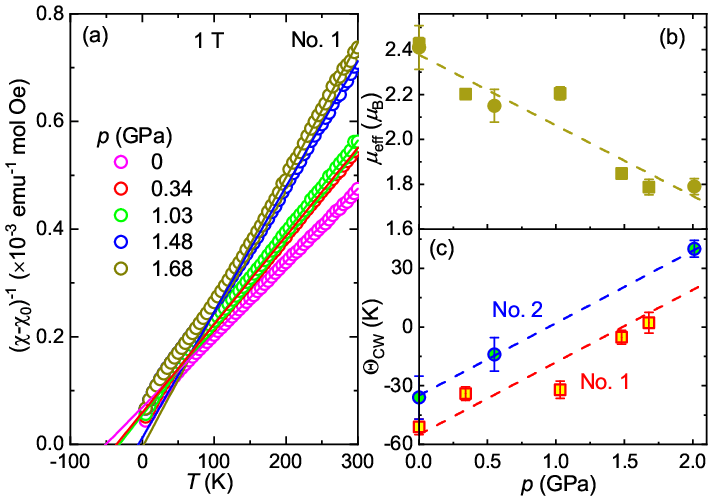}
	\vspace{-12pt} \caption{\label{Fig1} Curie-Weiss analysis of Li$_2$RhO$_3$ under pressure. (a) Temperature-dependent inverse dc magnetic susceptibility $H/M$ of Li$_2$RhO$_3$ measured at various pressures from 2~K to 300~K in a magnetic field of 1~T for run No. 1. Solid lines show the Curie-Weiss fit. Pressure evolution of (b) the effective moment $\mu_{\rm{eff}}$ and (c) the Curie-Weiss temperature $\Theta_{\rm{CW}}$ for run No. 1 (square symbols) and No. 2 (circle symbols). Dashed lines are linear fits.}
	\vspace{-12pt}
\end{figure*}

\begin{figure*}
	\includegraphics[angle=0,width=0.6\textwidth]{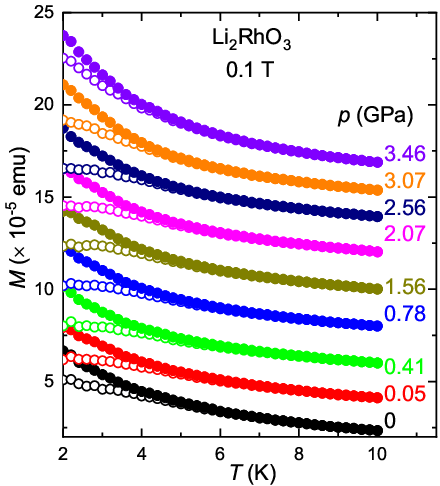}
	\vspace{-12pt} \caption{\label{Fig2} Temperature-dependent dc magnetic susceptibility $M(T)$ of Li$_2$RhO$_3$ measured at various pressures in a magnetic field of 0.1~T for run No. 3. The data are vertically offset for clarity. The solid symbols show the data collected upon field cooling (FC), whereas open symbols are the data collected upon zero-field cooling (ZFC).}
	\vspace{-12pt}
\end{figure*}

\begin{figure*}
	\includegraphics[angle=0,width=1\textwidth]{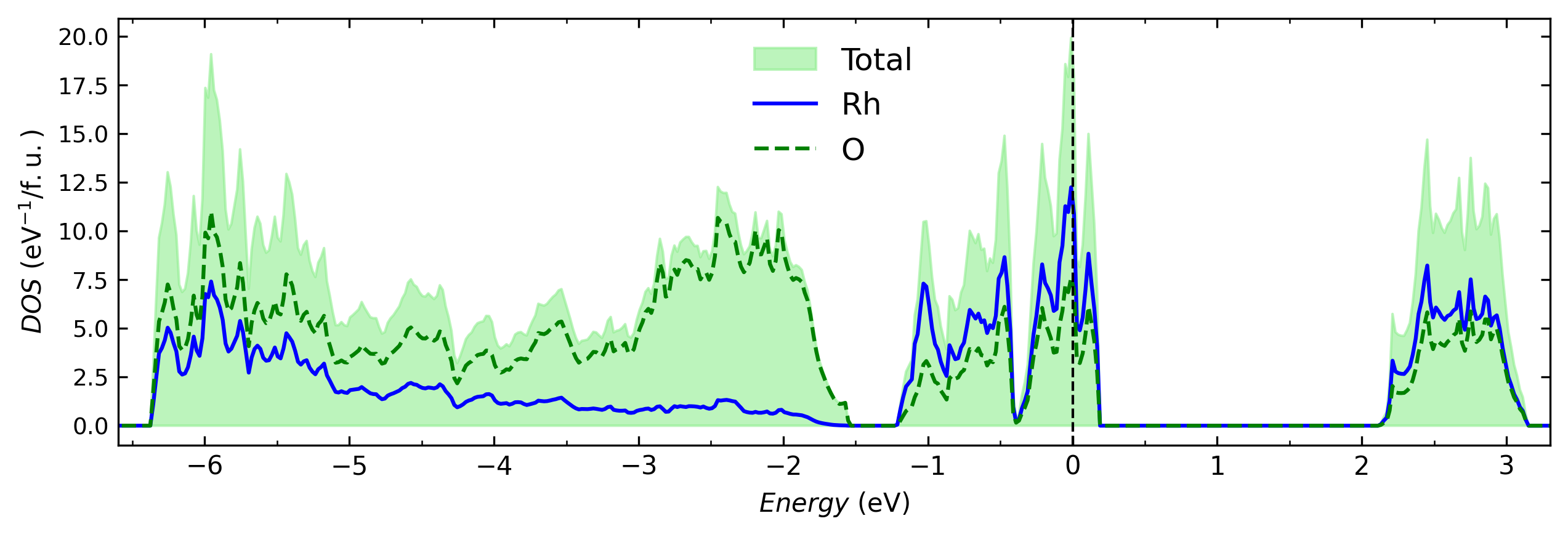}
	\vspace{-12pt} \caption{\label{PBE+SO} Full-relativistic (PBE+SO) density of states for \ce{Li2RhO3} calculated without spin polarization at ambient pressure. The Fermi level is at zero energy.
		%PBE and PBE+SO are very similar
	}
	\vspace{-12pt}
\end{figure*}

\begin{figure*}[h!]
	\includegraphics[angle=0,width=0.6\textwidth]{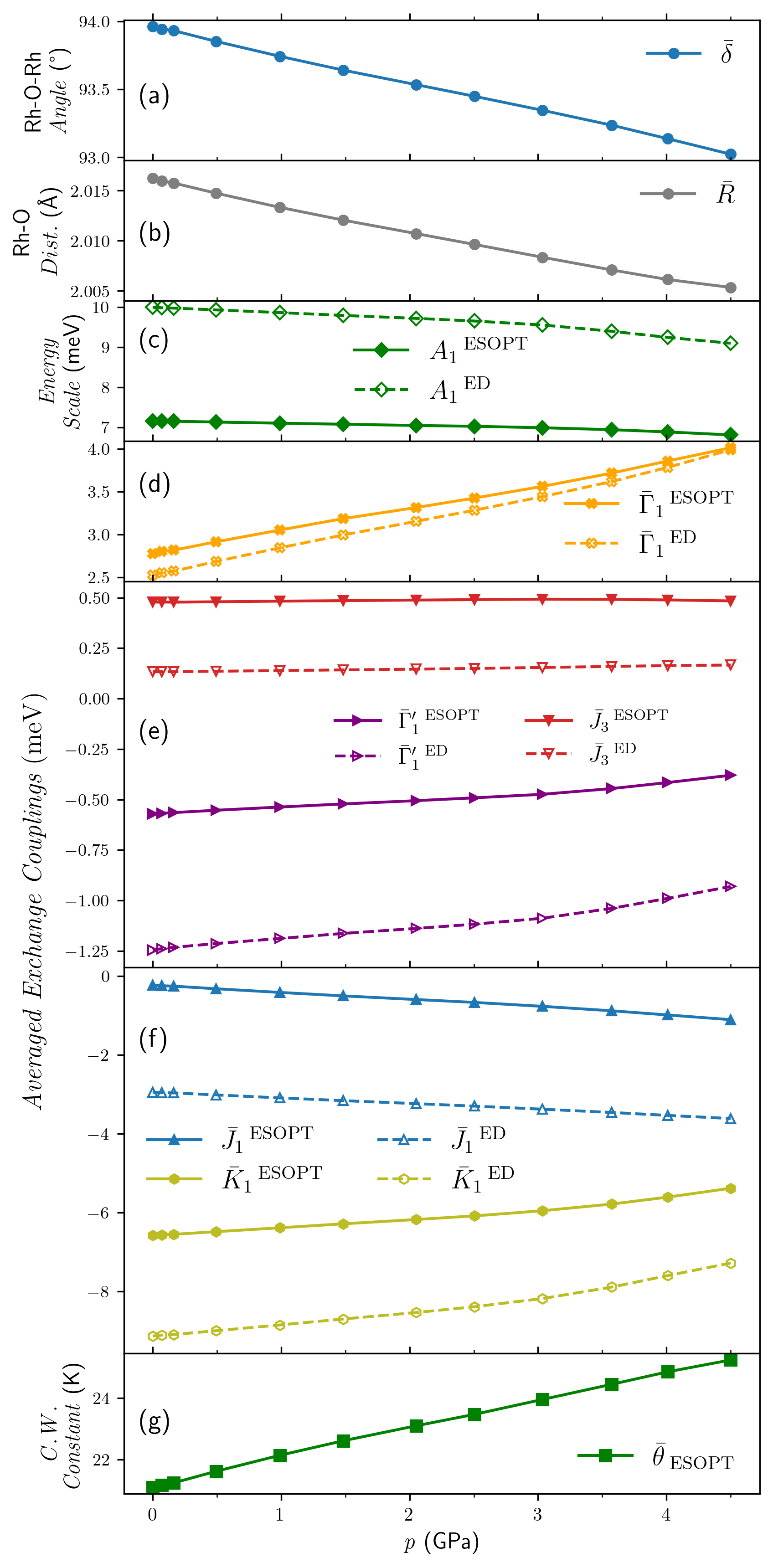}
	\vspace{-18pt} \caption{\label{JKGLRO} Pressure evolution of the averaged magnetic couplings in Li$_2$RhO$_3$ calculated by exact cluster diagonalization (ED) and second-order perturbation theory (ESOPT). (a) Averaged \ce{Rh-O-Rh} angle $\bar{\delta}$. (b) Averaged Rh-O distance, $\bar{R}$. (c) Overall energy scale $A=(\bar{J_1}^2 + \bar{K_1}^2 +\bar{\Gamma_1}^2 + \bar{\Gamma_1}^{'2})^{\frac12}$. (d) Off-diagonal anisotropy $\bar{\Gamma_1}$. (e) Off-diagonal anisotropy $\bar{\Gamma_1'}$ and the third-neighbor coupling $\bar{J_3}$. (f) Nearest-neighbor Kitaev ($\bar K_1$) and Heisenberg ($\bar J_1$) couplings. (g) Calculated pressure-dependent Curie-Weiss temperature. All lines are guides to the eye only.
	}
	\vspace{-18pt}
\end{figure*}

\begin{figure*}
	\includegraphics[angle=0,width=0.8\textwidth]{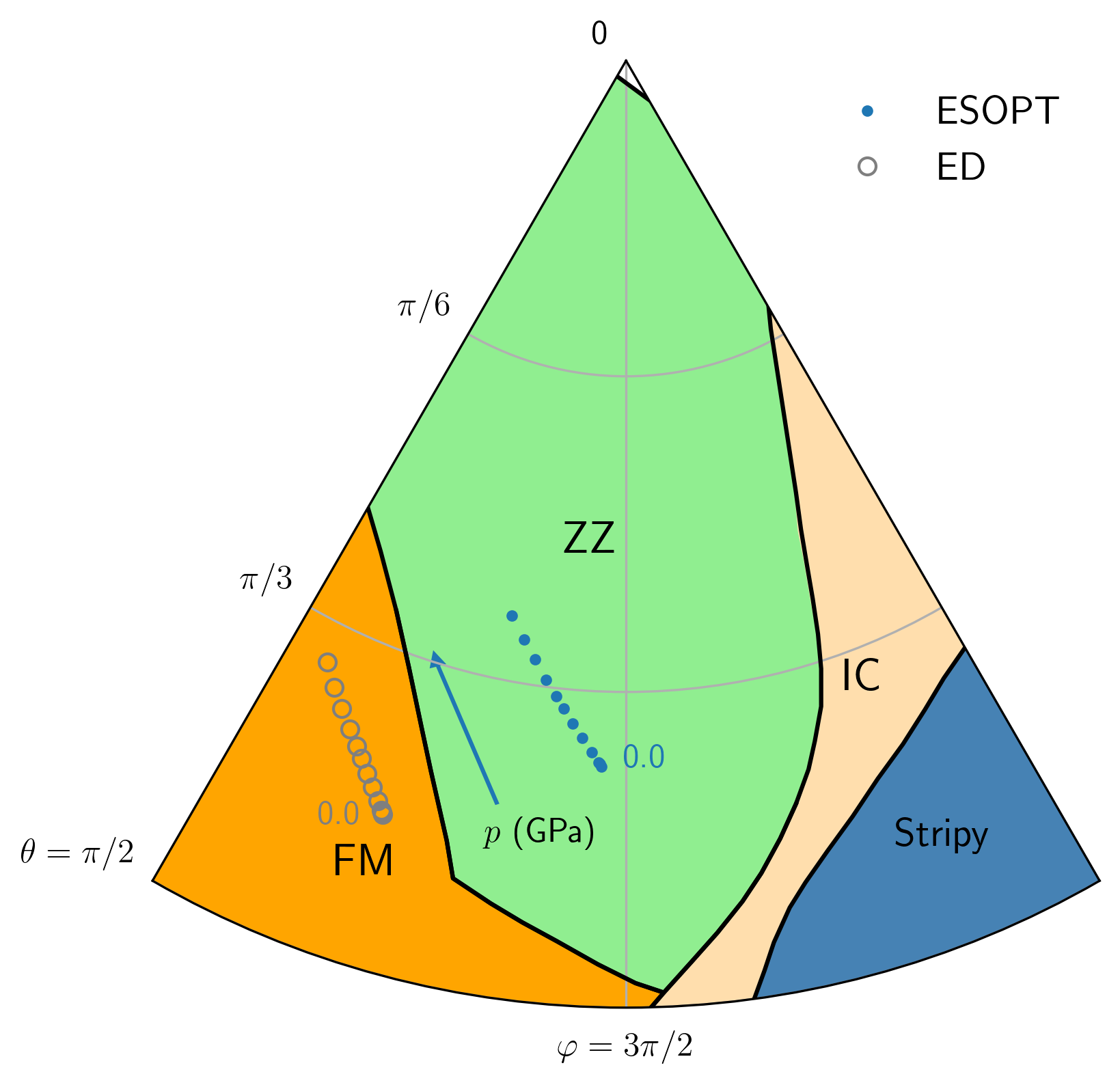}
	\vspace{-12pt} \caption{\label{classPD} \ce{Li2RhO3} placed onto the classical phase diagram of the $J_1-K_1-\Gamma_1$ model \cite{UnpubRau2014} with $\Gamma '/A =-0.10$. The parameterization in polar coordinates corresponds to $J_1/A=\sin{\theta}\cos{\varphi}$, $K_1/A=\sin{\theta}\sin{\varphi}$, and $\Gamma_1/A= \cos{\theta}$ where $\theta$ is the radial part, $\varphi$ is the angular part, and $A$ is the overall energy scale defined in the caption of Fig.~\ref{JKGLRO}. The values averaged over the $X$-, $Y$-, and $Z$-bonds are used. Note that $J_3$ is not included. It is expected to stabilize zigzag order (ZZ) at the expense of the ferromagnetic order (FM).
	}
	\vspace{-12pt}
\end{figure*}

	Density-functional band-structure calculations were performed in the FPLO code~\cite{fplo} using the Perdew-Burke-Ernzerhof (PBE) version of the exchange-correlation potential~\cite{pbe96}. Hopping parameters were obtained by the built-in Wannierization procedure of FPLO~\cite{koepernik2023}. Previous studies reported only the lattice parameters of Li$_2$RhO$_3$ as a function of pressure~\cite{19HerPRB}, whereas oxygen positions had a large uncertainty due to limitations of the x-ray powder diffraction data. Therefore, we chose to fix the lattice parameters to their experimental values at each pressure and relaxed the atomic positions by minimizing forces on atoms within FPLO. \footnote{DFT+$U$ calculations were used for structural relaxation with the residual forces of less than $\SI{1e-3}{\electronvolt\per\angstrom}$.}
	At ambient pressure, we obtained the Rh--O distances ($R_1, R_2, R_3$ in Fig.~\ref{CrisSt}(a)) of 2.0395/1.9713/2.0379\,\r A, the Rh--Rh distances ($Z, XY$) of 2.9812/2.9315\,\r A, and the Rh--O--Rh bond angles (labeled as $\delta_1, \delta_2$) of 93.91/94.01$^{\circ}$, which are in a good agreement with the experimental values~\cite{11TodZAAC} of 2.033(6)/2.001(7)/2.029(3)/\,\r A  (Rh--O1/Rh--O1/Rh--O2), 2.950(4)/2.954(5)\,\r A (Rh--Rh), and 94.16/93.29$^{\circ}$ (Rh--O1--Rh/Rh--O2--Rh). We also note that the experimental structural data for Li$_2$RhO$_3$ feature a weak Li/Rh site mixing~\cite{11TodZAAC}. This site mixing is caused by the interlayer disorder, such as stacking faults, whereas individual honeycomb layers are well-ordered~\cite{bahrami2022}. Therefore, we used the fully ordered structural model in our calculations.
	
	Magnetic couplings in Li$_2$RhO$_3$ are defined by the general spin Hamiltonian,
	\begin{equation}
		\mathcal H=\sum_{\langle ij\rangle} \mathbf S_i\mathbb J_{ij}\mathbf S_j
	\end{equation}
	where $\mathbb J_{ij}$ is the exchange tensor for the respective bond, and the summation is over bonds. The $\mathbb J_{ij}$ components were determined by two complementary approaches. In the superexchange model, the hoppings within the $t_{2g}$ manifold of the scalar-relativistic band structure are used to calculate the exchange couplings as described in Ref.~\cite{winter2016}. Weak crystal-field splittings within the $t_{2g}$ manifold and virtual processes involving the $e_g$ states are neglected in this method.
	
	Cluster many-body calculations allow a comprehensive treatment of the microscopic processes that underlie the exchange couplings. In order to calculate exchange couplings, we start by deriving the electronic Hamiltonian in terms of Rh 4\textit{d} orbitals Wannier basis~\cite{koepernik2023}. We obtain this Hamiltoninan by performing fully relativistic PBE calculations using a \textit{k}- grid of $12\times 12 \times 12$ and retaining the translational symmetry of the system. The obtained electronic Hamiltonian is exactly diagonalized on two-site clusters to get the low-energy eigenstates. These eigenstates are then projected to pure spin states by following the des Cloizeaux effective Hamiltonian method~\cite{des1960extension} to obtain the intermediate states which are finally orthonormalized by employing the symmetric (L\"owden) approach \cite{lowdin1950non}. The advantage of the above described procedure is that it preserves all the symmetries and includes the effects of the non-cubic crystal-field splitting and nominally empty $e_g$ orbitals that are neglected in the superexchange model.                    
	
	In both cases, we used the same parameters of $U=2.58$\,eV and $J_H=0.29$\,eV for the on-site Coulomb repulsion and Hund's coupling, respectively, as determined for $\alpha$-RuCl$_3$~\cite{eichstaedt2019deriving}. The spin-orbit coupling $\lambda=0.15$\,eV was used in the superexchange model.

	%-------------------------------------------------------------------------------------------------------
	\section{Data Availability}
	The experimental and computational data associated with this manuscript are available from Ref.~\cite{DATA}.
	
	\section{Code availability}
	All codes in this paper are available from the corresponding authors upon reasonable request.
	
	\section{Acknowledgments}
	This work was funded by the Deutsche Forschungsgemeinschaft (DFG, German Research Foundation) -- TRR 360 -- 492547816 (subproject B1). B.S. acknowledges the financial support of Alexander von Humboldt Foundation.
	
	\section{Author contributions}
	B.S. carried out the magnetization measurement under pressure. E.I., A.-A.T., R.D. and S.-M.W. performed the ab initio and cluster many-body calculations. F.F. synthesized the powder of Li$_2$RhO$_3$. P.G. and A.-A.T. designed the project. All authors contributed to analyzing the data, discussions, and the writing of the manuscript.
	\section{Competing Interests}
	The authors declare no competing interests.

\widetext
	\clearpage
	\begin{center}
		\textbf{Supplementary Information: Pressure-dependent Magnetism of the Kitaev Candidate Li$_2$RhO$_3$}
	\end{center}
	%%%%%%%%%%% Merge with supplemental materials %%%%%%%%%%
	%%%%%%%%%%% Prefix a "S" to all equations, figures, tables and reset the counter %%%%%%%%%%
	\setcounter{equation}{0}
	\setcounter{figure}{0}
	\setcounter{table}{0}
	\setcounter{page}{1}
	\makeatletter
	\setcounter{section}{0}
	\renewcommand{\thesection}{S-\Roman{section}}
	\renewcommand{\thetable}{S\arabic{table}}
	\renewcommand{\theequation}{S\arabic{equation}}
	\renewcommand{\thefigure}{S\arabic{figure}}
	\renewcommand{\bibnumfmt}[1]{[S#1]}
	\renewcommand{\citenumfont}[1]{S#1}
    
    %\section{Complete magnetic couplings under pressure}

\begin{figure}[htbp]
\centering
\includegraphics[width=1\linewidth]{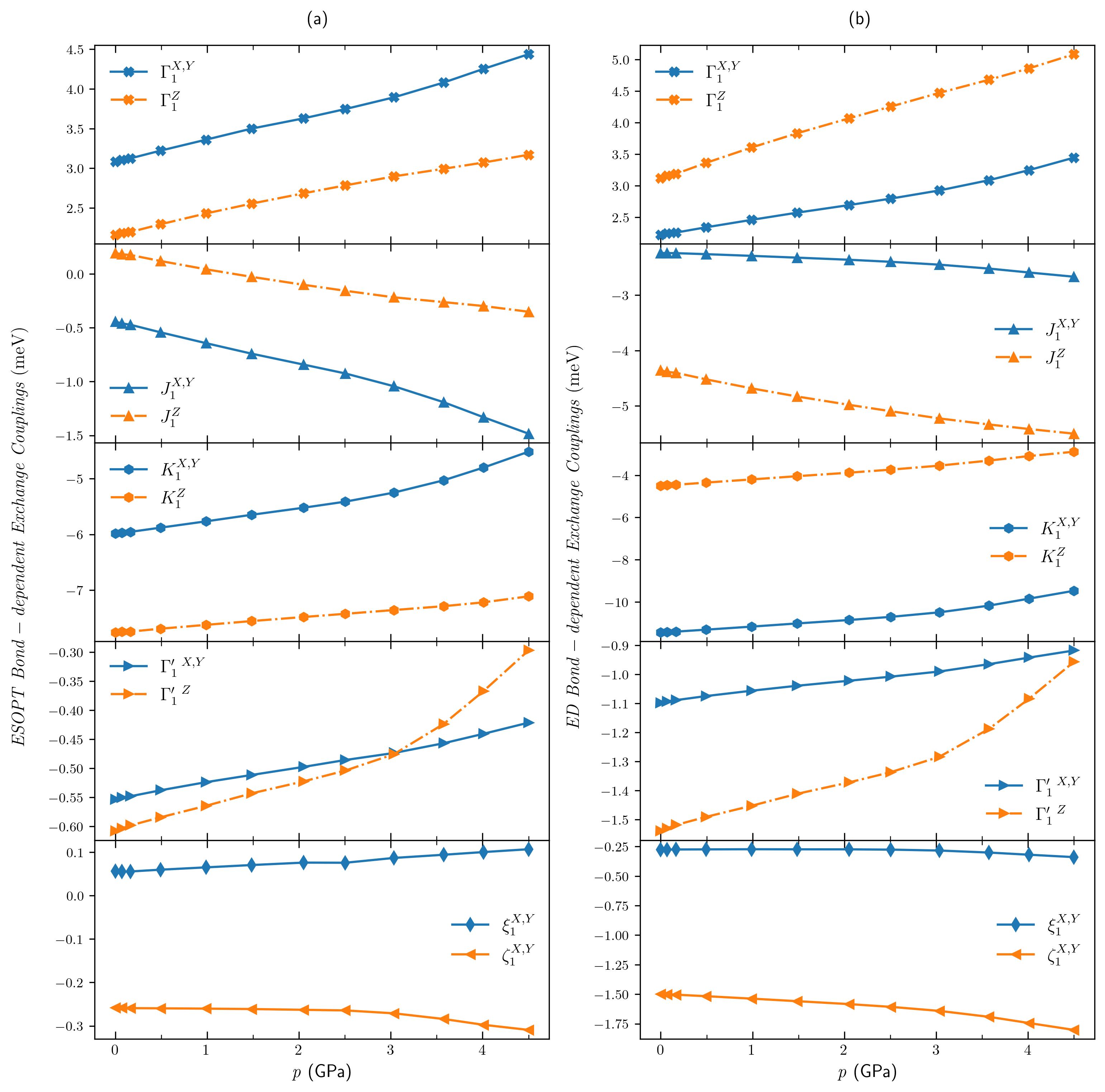}
\caption{Complete set of magnetic exchange parameters for nearest-neighbor interactions in Li$_2$RhO$_3$. (a) Exact second-order perturbation theory (ESOPT). (b) Exact diagonalization (ED) for two-site clusters. All lines are guides to the eyes only.}
\label{fig:ED-ESOPT}
\end{figure}

In addition to equation (2) of the main text, the lower symmetry bonds $X$ and $Y$ are parametrized by equations \ref{eq:Jx} and \ref{eq:Jy} respectively \cite{winter2016}:

\begin{equation}
\mathbb J_n^X=\begin{pmatrix}
    J_n^{X}+K_n^{X} & \Gamma _n’^{X}+ \zeta_n &\Gamma _n’^{X}- \zeta_n \\
    \Gamma _n’^{X}+ \zeta_n & J_n^{X}+\xi_n & \Gamma _n^{X} \\
    \Gamma _n’^{X}-\zeta_n& \Gamma _n^{X} &J_n^{X} -\xi_n^{X}
\end{pmatrix}
\label{eq:Jx}
\end{equation}

\begin{equation}
\mathbb J_n^Y=\begin{pmatrix}
    J_n^{Y} + \xi_n & \Gamma_n’^{Y} + \zeta_n & \Gamma_n^{Y} \\
    \Gamma_n’^{Y} + \zeta_n & J_n^{Y} + K_n^{Y} & \Gamma_n’^{Y} - \zeta_n \\
    \Gamma_n^{Y}  & \Gamma_n’^{Y} - \zeta_n&J_n^{Y} -\xi_n´
\end{pmatrix}
\label{eq:Jy}
\end{equation}

which reflect the trigonal distortion of the octahedral cage surrounding the \ce{Rh} atom for the $n$ nearest neighbors. The magnetic constants for the X$_n$ and Y$_n$ bonds turn out to be the same.
Based on these parametrizations we obtain magnetic couplings using the superexchange theory (ESOPT) and the exact cluster diagonalization (ED) method for first and third nearest neighbors, as shown in figures \ref{fig:ED-ESOPT} and \ref{fig:ED-ESOPT-3nn} respectively.

\begin{figure}[htbp]
\centering
\includegraphics[width=1\linewidth]{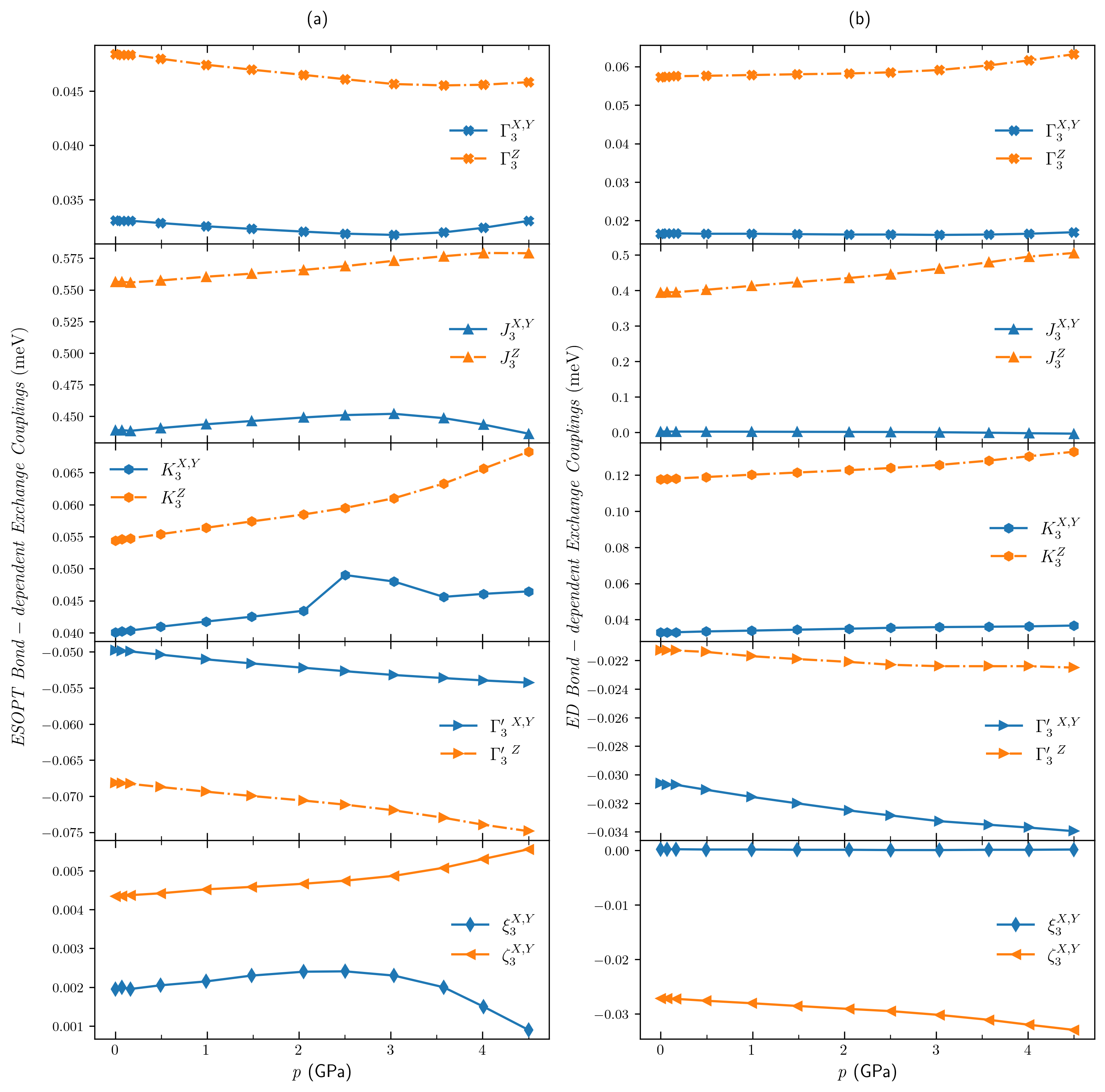}
\caption{Complete set of magnetic exchange parameters for third-neighbor interactions in Li$_2$RhO$_3$. (a) Exact second-order perturbation theory (ESOPT). (b) Exact diagonalization (ED) for two-site clusters. All lines are guides to the eyes only.}
\label{fig:ED-ESOPT-3nn}
\end{figure}

%\section{Rh-O-Rh angles and Rh-O distances from force optimization}

\begin{figure}[htbp]
\centering
\includegraphics[width=1\linewidth]{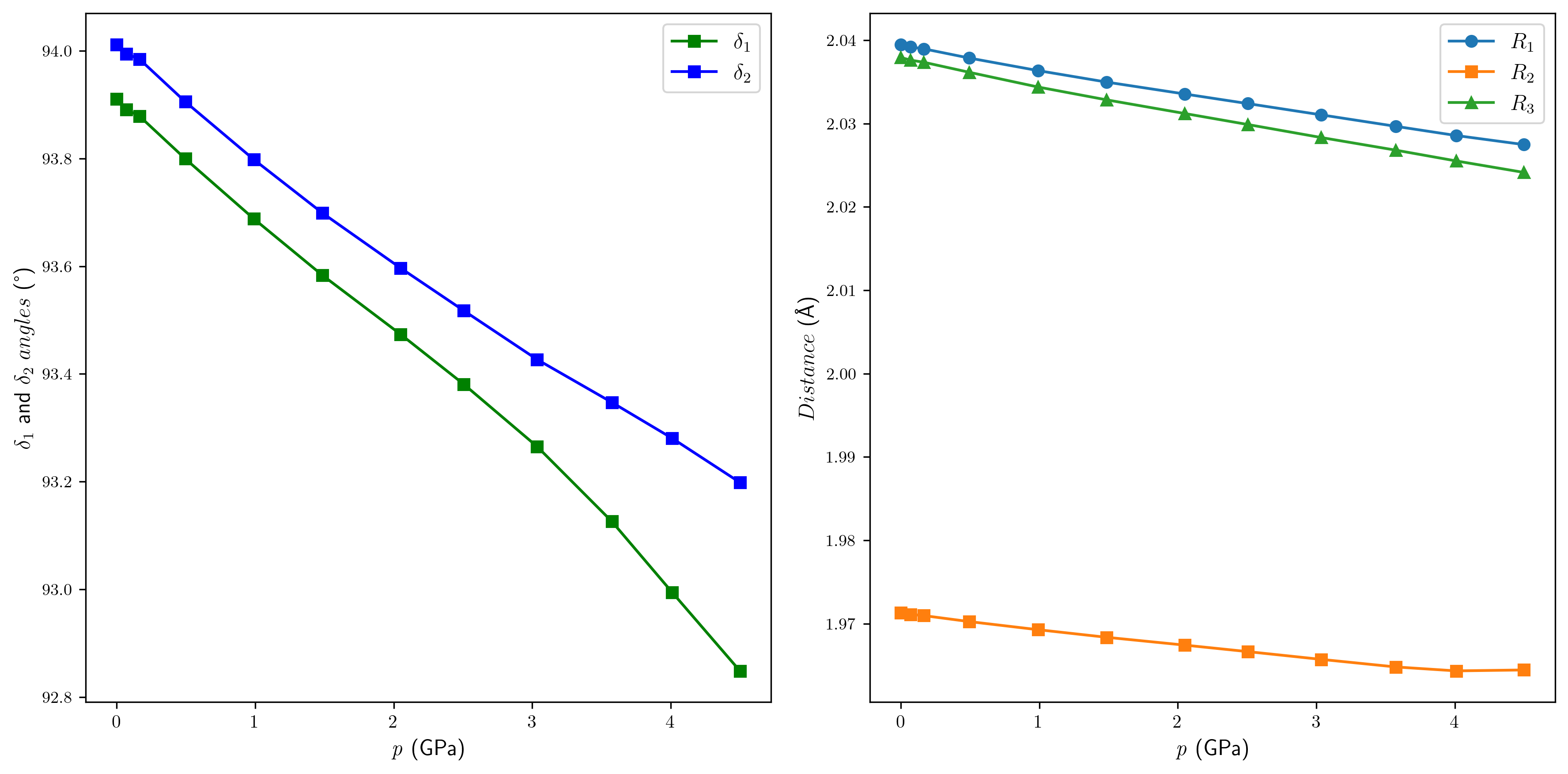}
\caption{Results of the force optimization for (a) the Rh-O-Rh angles ($\delta_1$ and $\delta_2$) and (b) the Rh-O distances ($R_1, R_2, R_3$). All lines are guides to the eyes only.}
\label{fig:ED-ESOPT-3nn}
\end{figure}

% End single-column mode

\onecolumngrid

\end{document}